\documentclass[a4paper]{cas-dc}

\usepackage[numbers,sort&compress]{natbib}
\usepackage{booktabs}
\usepackage{multirow}
\usepackage{amssymb}
\usepackage{threeparttable}
\usepackage{bigstrut}
\usepackage{multirow}
\usepackage{float}
\usepackage[misc]{ifsym}
\usepackage{algorithmic}
\usepackage{algorithm}
\usepackage{pifont}
\usepackage{caption} 
\def\tsc#1{\csdef{#1}{\textsc{\lowercase{#1}}\xspace}}
\tsc{WGM}
\tsc{QE}
\tsc{EP}
\tsc{PMS}
\tsc{BEC}
\tsc{DE}

\begin{document}
	\begin{sloppypar}
		\let\WriteBookmarks\relax
		\def\floatpagepagefraction{1}
		\def\textpagefraction{.001}
		\shorttitle{Leveraging social media news}
		\shortauthors{C. Wang et~al.}
		
		\title [mode = title]{CGCCE-Net:Change-Guided Cross Correlation Enhancement Network for Remote Sensing Building Change Detection
		}

		\author[1,3]{Chengming Wang}[style=chinese,
		type=editor,
		auid=000,bioid=1,
		role=Researcher,
		]
		\ead{2023420195@sdtbu.edu.cn}
		
		\address[1]{School of Computer Science and Technology, Shandong Technology and Business University, Yantai 264005, China}

		\begin{abstract}
			Change detection encompasses a variety of task types, and the goal of building change detection (BCD) tasks is to accurately locate buildings and distinguish changed building areas. In recent years, various deep learning-based BCD methods have achieved significant success in detecting difference regions by using different change information enhancement techniques, effectively improving the precision of BCD tasks. To address the issue of BCD with special colors, we propose the change-guided cross correlation enhancement network (CGCCE-Net). We design the change-guided residual refinement (CGRR) Branch, which focuses on extending shallow texture features to multiple scale features obtained from PVT, enabling early attention and acquisition of special colors. Then, channel spatial attention is used in the deep features to achieve independent information enhancement. Additionally, we construct the global cross correlation module (GCCM) to facilitate semantic information interaction between bi-temporal images, establishing building and target recognition relationships between different images. Further semantic feature enhancement is achieved through the semantic cognitive enhancement module (SCEM), and finally, the cross fusion decoder (CFD) is used for change information fusion and image reconstruction. Extensive experiments on three public datasets demonstrate that our CGCCE-Net outperforms mainstream BCD methods with outstanding performance.
		\end{abstract}
		
		
		\begin{highlights}
			\item The change-guided residual refinement branch is proposed.
			\item Global cross correlation module for establishing semantic relations in bi-temporal images proposed.
			\item The semantic cognitive enhancement module provides enhancements for advanced semantic interaction.
		\end{highlights}
		
		
		\maketitle
		
		\section{Introduction}
		Change detection aims to monitor the spatiotemporal changes of ground objects by comparing bi-temporal remote sensing images of the same area taken at different times. This technology plays a significant role in monitoring urban development, environmental changes, and other factors, helping administrators understand land use, urban expansion, and other key elements from a broader perspective. With the application of high-resolution bi-temporal remote sensing images and the continuous development of sensor technology, change detection has been widely applied in various fields such as disaster monitoring, land management, and environmental protection \cite{1,2,3,4}. For example, by comparing remote sensing images taken at different times, significant changes in urban areas can be identified, including land use changes, the construction and demolition of buildings, as well as vegetation cover changes caused by natural disasters. In addition to global factors, geographical features of the region itself, such as seasonal variations in vegetation cover and roof color differences, can also interfere with detection results. For instance, seasonal changes may cause objects that should remain stable to exhibit significant color variations at different time points. Furthermore, some real changes may be overlooked due to noise or lighting differences in the images. Therefore, the core challenge of change detection lies in accurately distinguishing real changes from these false changes, ensuring the correct identification of change targets.
		
		\begin{figure}[!t]
			\centering
			\includegraphics[width=3.5in]{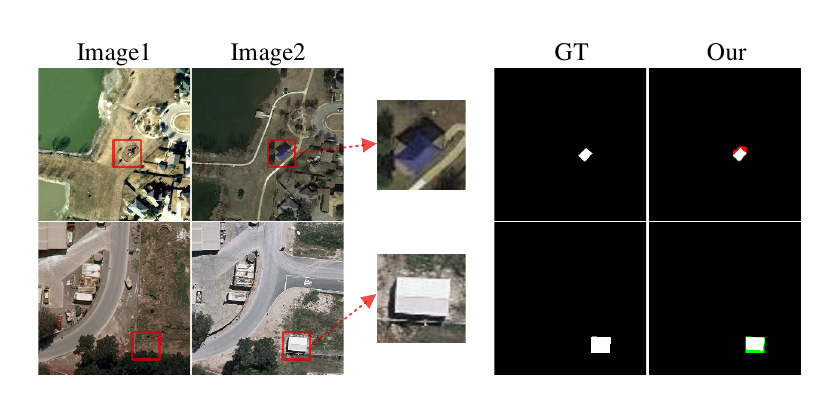}
			\caption{The demonstration highlights the problem of building change detection with special colors, including buildings with deep purple and bright white roofs. The specific visual comparison results are shown in Fig. \ref{fig8}.
			}
			\label{fig1}
			\vspace{-15pt}
		\end{figure}
		
		Traditional BCD methods mainly include pixel-based analysis methods and feature-based analysis methods. Among pixel-based analysis methods, the most common ones are image differencing and principal component analysis (PCA) \cite{5,6,7}. Image differencing identifies changes by calculating the pixel differences between bi-temporal images, but it is easily affected by lighting and seasonal variations. PCA, on the other hand, reduces the dimensionality of image data and extracts the main change information to reduce redundancy. However, its limitation lies in its primary applicability to linear changes, making it less effective in handling complex nonlinear changes. Additionally, feature-based analysis methods detect changes by extracting features such as texture, color, and shape from images\cite{8,9}. For example, texture-based BCD methods distinguish land cover changes by analyzing texture features of the images. When facing dynamic environments and complex data, machine learning methods can more effectively handle high-dimensional data and nonlinear changes. These methods automatically learn and extract key features from the data by training models. Common machine learning methods include support vector machines (SVM) \cite{10,11} and random forests (RF) \cite{12,13}. SVM is a supervised learning method that constructs a decision hyperplane to differentiate between changed and unchanged areas. SVM is particularly suitable for handling high-dimensional data and can effectively address nonlinear changes, avoiding the shortcomings of traditional methods that overly rely on linear features. RF, as an ensemble learning method, constructs multiple decision trees and uses voting to determine the change category. It can automatically process a large number of features and effectively reduce the overfitting problem.

		With the rapid iteration of hardware facilities and the deep understanding of algorithms, the development and application of deep learning technologies have gained wide spread attention. Fully convolutional network (FCN) \cite{14,15} focus on pixel-level image segmentation by replacing traditional fully connected layers with convolutional layers, enabling pixel-level classification predictions of input images. Convolutional Neural Network (CNN) use convolutional kernels to compute hierarchical features of images, excelling in local feature extraction. The Transformer \cite{16} is a network architecture centered around the self-attention mechanism, which effectively processes sequential data and captures long-distance dependencies through global information modeling. When handling bi-temporal images, it overcomes the limitations of traditional CNN in long-distance modeling by linking global relationships and constructing semantic information. The attention mechanism assigns different weights to different parts of the input, allowing the network to focus on important regions of the image and enhancing the model's ability to capture critical information in change regions. Among these, the spatial attention mechanism assigns different weights to different spatial positions of the image, helping the model identify subtle changes between different land cover types \cite{17,18}, while channel attention assigns different weights to channels, allowing the model to dynamically adjust between feature maps \cite{19,20}. The combination of these two attention mechanisms further highlights critical information. Numerous methods combining attention mechanisms with CNN and Transformers have emerged, and using network architectures in conjunction with attention mechanisms at different training stages can effectively address various problems in corresponding tasks \cite{49,50,51,52,53}.
		
		However, despite the success of novel methods proposed for BCD tasks, they still have limitations when handling specific issues. The main task in remote sensing change detection is to detect building changes in two remote sensing images of the same area at different times. The specific issues we identify include three categories: extensions or incomplete demolitions of existing buildings, tree obstructions to buildings, and change detection of buildings with special colors. Among these, the failure of detecting buildings with special colors has drawn our attention, as shown in Fig. \ref{fig1}. We analyze two possible reasons for this problem from the model's perspective: one is that the model has only learned to detect changes in buildings with similar colors, and the other is that the model classifies such buildings as other environmental features, rather than as buildings. Therefore, we aim to improve the detection of buildings with special colors by focusing on two aspects: guiding change information based on local textures and semantic interaction of bi-temporal images.

		\begin{figure*}
			\centering
			\includegraphics[width=1\textwidth]{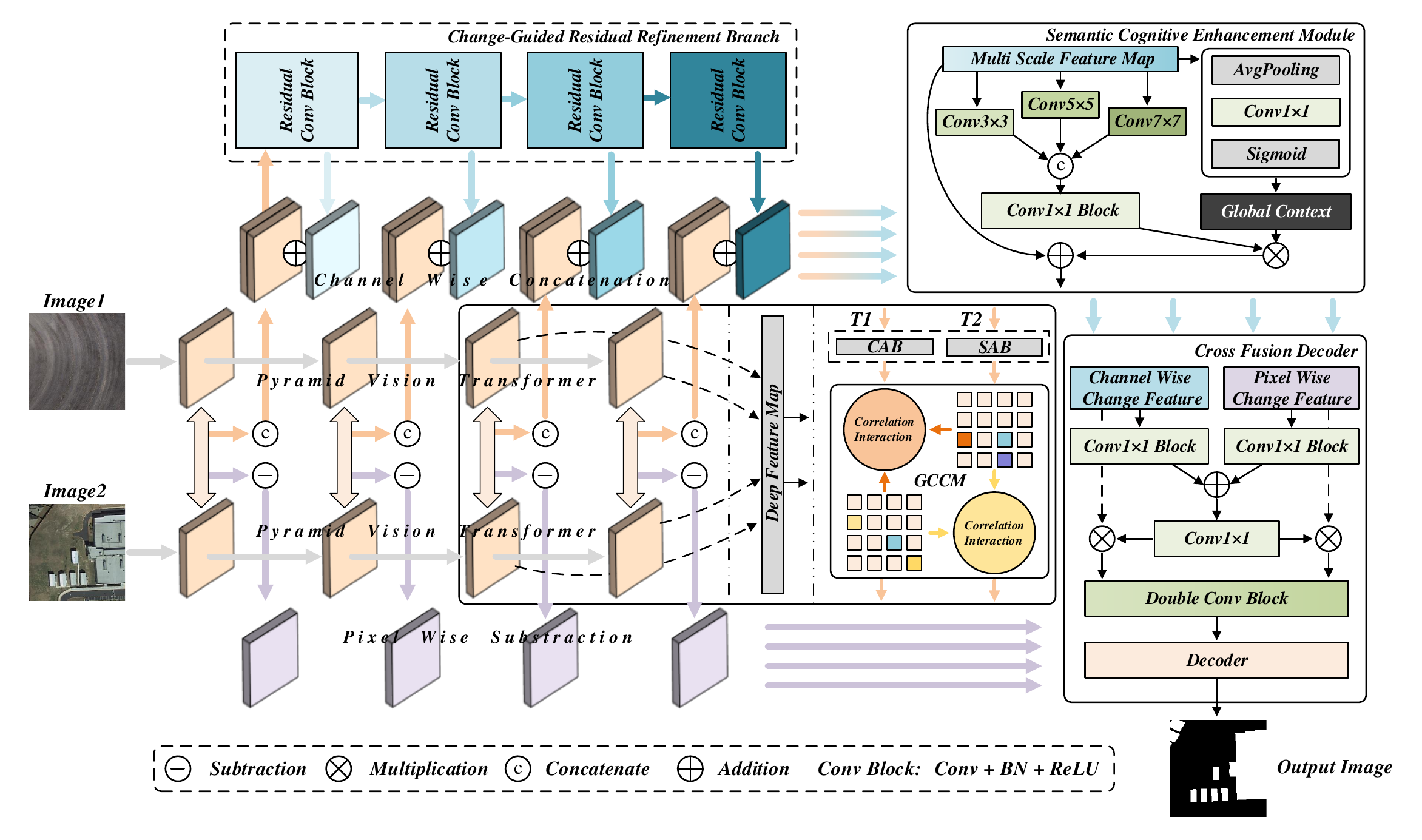}
			\caption{The overall architecture of CGCCE-Net is shown. It includes the PVT encoder that generates multi-scale features, the CGRR branch for extracting local texture information, the GCCM for semantic information interaction between bi-temporal images, the SCEM for enhancing semantic cognition, and the CFD for change information fusion and image reconstruction.
			}
			\label{fig2}
			\vspace{-10pt}
		\end{figure*}
		To address the above issues, we propose the CGCCE-Net. First, we use the pyramid vision transformer (PVT) \cite{21,22} as a dual-branch architecture to process the bi-temporal images separately. The feature maps generated by the two branches are then concatenated at the channel level and differenced at the pixel level to obtain two types of feature maps containing change information. To avoid the model overfitting to the detection of changes in buildings with similar colors, we design a branch based on a residual refinement structure to guide the change information of the original branch. This branch takes the shallow features from PVT as input and is designed with multi-stage outputs to adapt to a multi-level pyramid structure, ensuring that both shallow and deep features are deeply integrated with the change information containing texture and color features. Additionally, to enhance the model's semantic understanding of the bi-temporal images, we design the GCCM and SCEM modules, enabling the mutual recognition of buildings and changes between the two branches. Finally, to further strengthen the stability of change detection and effectively preserve spatial information, we design the CFD to perform image reconstruction. The main contributions of this paper are as follows:
		
		$\bullet$ We designed CGCCE-Net for remote sensing building change detection (RSCD) tasks, using the PVT encoder to obtain shallow features, and passing these features containing texture information through the CGRR branch to guide early change information for the original branch.
		
		$\bullet$ We designed GCCM and SCEM to enhance the model's semantic understanding capability. GCCM establishes the interaction of global and semantic information between the bi-temporal images, while SCEM integrates the texture change information from the CGRR branch and the high-level semantic information from GCCM, achieving enhanced semantic understanding in the model.
		
		$\bullet$ Extensive experiments on three public datasets confirm that CGCCE-Net outperforms other BCD methods, demonstrating superior performance.

		\section{Related work}
		In the existing RSCD tasks, there are two main approaches based on deep learning: one is CNN-based methods, and the other is Transformer-based methods. Additionally, novel attention mechanisms have also been widely applied in BCD tasks.
		\subsection{BCD Methods based on CNN}
		Many existing BCD methods have referenced models designed for image segmentation tasks, such as U-Net \cite{23,24}. U-Net achieves significant success by utilizing a classic U-shaped architecture with skip connection mechanisms, providing a simple and efficient flexible structure. FC-SC, FC-SD, and FC-EF \cite{25} are all built on the U-Net architecture for BCD tasks, using three different full convolutional network structures to achieve concatenation, differencing, and fusion, marking the early exploration of BCD methods. STANet \cite{26} proposes the Spatiotemporal Adaptive Network to preserve texture information in image reconstruction while ensuring the model's ability to capture global information, aiming to refine temporal features and transfer spatial textures. IFNet \cite{27}, to improve the edge accuracy of change regions and eliminate false change interference, adopts a deep supervision mechanism, calculating loss with multiple losses jointly to optimize the model. SNUNet \cite{28} reduces the uncertainty of target edge pixels and the occurrence of small target detection failures by using densely connected siamese networks.
		
		CNN-based BCD methods have performed well in RSCD tasks, excelling at local feature extraction. However, CNNs struggle to effectively capture global information in bi-temporal remote sensing images. Therefore, we leverage the advantage of CNNs in extracting local information and construct a CNN-based residual block to extract local texture features for early change information guidance in the original branch.
		
		\subsection{BCD Methods based on Attention Mechanism}
		In RSCD tasks, global noise includes factors like lighting, seasonal, and color changes, while local noise includes features such as forests and non-target buildings. Under the influence of multiple types of noise, the combination of attention mechanisms with the model becomes particularly important. The attention mechanism assigns different weights, focusing the model's attention on the target region, thus helping the model focus on the BCD task. STADE-CDNet \cite{29} introduces a spatiotemporal attention mechanism to enhance differences in the network. By using the temporal memory module (TMM), it extracts temporal and spatial information, addressing class imbalance issues and false change interference caused by lighting changes. SAAN \cite{30} designs similarity-aware attention to guide the similarity optimization in the deep encoder layers, enhancing the semantic relationships of bi-temporal images and solving the problem of missing explicit semantic information. DAMFA-Net \cite{31} proposes using a dual-attention fusion module during the decoding stage to guide the fusion of features at different scales, solving the loss of edge details and small target features. ELGC-Net \cite{32} introduces an efficient local-global context module that uses novel pooling-transpose attention to capture enhanced global context for accurate change detection. In the application of general attention mechanisms, the linear angle attention mechanism proposed in Castling-ViT \cite{33} has provided significant inspiration. 
		
		Many novel attention mechanisms combined with BCD methods have achieved great success in constructing models, indicating that a suitable attention mechanism can significantly improve model performance. However, the use of attention mechanisms in RSCD often lacks specificity. We use linear attention to build the GCCM to realize global information interaction and semantic information enhancement of bi-temporal images in the deep features.
		
		\begin{figure}[!t]
			\centering
			\includegraphics[width=3.5in]{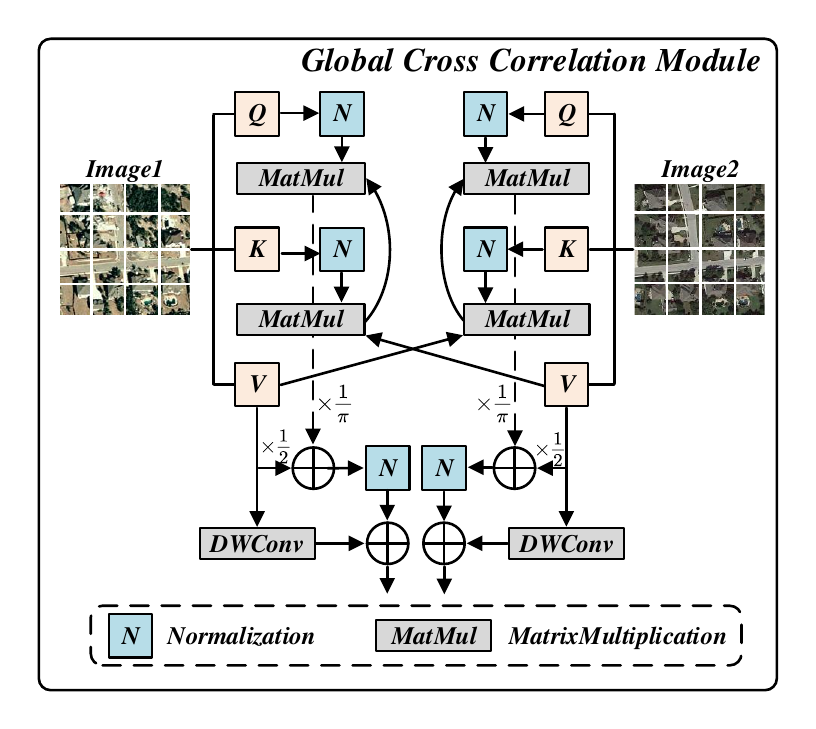}
			\caption{The demonstration illustrates the basic architecture of GCCM. Based on the unique task of RSCD, we have specifically designed GCCM for semantic information interaction between bi-temporal images using linear angle attention.}
			\label{fig3}
			\vspace{-10pt}
		\end{figure}
		
		\subsection{BCD Methods based on Transformer}
		Transformer has a powerful self-attention mechanism, which is excellent at processing sequential data and capturing long-range dependencies between data in natural language processing. The Vision Transformer (ViT) \cite{34} extends this ability to acquire global information in image processing, allowing the Transformer architecture to thrive in computer vision and be widely applied in RSCD tasks. BIT \cite{35} introduces Transformer to better model contextual information, marking an early application of Transformer in BCD tasks and laying the foundation for the further development of Transformer architectures in BCD methods. GCFormer \cite{36} combines CNN and Transformer with a multi-receptive field mechanism to extract rich contextual information, and proposes relative position encoding to replace the absolute position encoding in Transformer, effectively capturing long-range dependencies. STCD \cite{37}, based on the Swin Transformer \cite{38}, proposes a general BCD model without CNN, using a deconvolutional layer with axial attention for image reconstruction. DiFormer \cite{39} introduces a token-swapping-based difference evaluation module to generate inconsistency information at the boundary of the target region, thereby highlighting the target area.
		
		Transformer-based BCD methods have achieved significant success, primarily addressing issues such as class imbalance, small target detection, and edge detection ambiguity. Therefore, we use PVT as the core architecture of our model and implement multi-scale feature extraction of bi-temporal images through weight-sharing in a siamese network.

		\section{Methodology}
		In this section, we first introduce the overall structure of the model, followed by an overview of the PVT encoder. Next, we present the CGRR branch, followed by explanations of the GCCM and SCEM. Finally, we give an overview of the CFD for image reconstruction.
		
		\subsection{Motivation}
		Our model targets the problem of detecting changes in buildings with special colors in RSCD. The key challenge lies in how to effectively extract and apply local texture information and how to achieve semantic information interaction between the bi-temporal images. Specifically, the design motivation for the CGRR branch is to extract local texture information, aiming to address the BCD problem of buildings with special colors at the local information level. The design motivation for the GCCM is to achieve semantic cognitive interaction between the bi-temporal images, focusing on resolving the BCD problem of buildings with special colors at the global information and cognitive relationship level. The design motivation for the SCEM is to further enhance the local texture information extracted by the CGRR branch and the high-level semantic information generated by the GCCM. Its role is to effectively fuse the enhanced information and ensure the stability and coherence of the model when absorbing useful feature information. The design motivation for the CFD is to achieve the organic integration of change information, aiming to fuse the change information after multiple layers of processing by the siamese network and perform the image reconstruction process.
		
		\subsection{Overview}
		Fig. \ref{fig2} illustrates the overall architecture of our CGCCE-Net. We introduce the PVT encoder to process the bi-temporal images simultaneously with weight sharing, thereby obtaining multi-scale features. The introduction of the PVT encoder is aimed at addressing the inherent challenges in the RSCD task, including global and local pseudo-change interference. Based on this, we perform channel-level concatenation and pixel-level differencing on the multi-scale features obtained from the twin network, making the model more aligned with the BCD task. The raw features after encoding have not yet undergone spatiotemporal information interaction, so the model’s focus is on buildings. To enable early interaction of change information, we apply channel attention and spatial attention to independently enhance the original features. Subsequently, we design the GCCM, which is suitable for the RSCD task based on linear attention, to achieve high-level semantic information interaction of bi-temporal images at the raw feature stage. After analyzing the essence of the detection failure issue for buildings with special colors, we pinpoint the differences in the local texture information of the feature maps. Therefore, we select low-level features containing local texture information as prior knowledge and construct the CGRR branch. The independent training results of this branch are then fed back into the channel-level concatenated feature maps. The SCEM module is designed to effectively fuse the information from the CGRR branch while further enhancing semantic information across different scales. It serves as a crucial link between the encoding and decoding stages of the model. Finally, we structure the CFD to fuse the change information of the bi-temporal images and complete the image reconstruction.

		\begin{figure}[!t]
			\centering
			\includegraphics[width=3.5in]{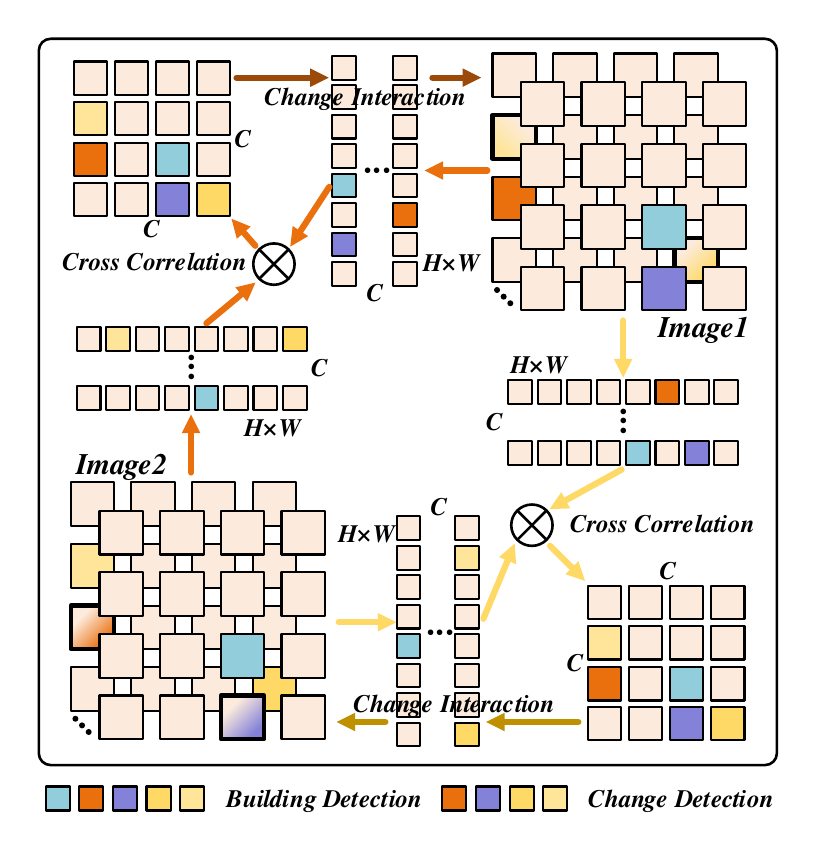}
			\caption{Detailed description of the cross-correlation mechanism of GCCM.After completing building detection, the bi-temporal images undergo cross-correlation to achieve semantic information interaction. The deepened black box can be abstractly represented as the change information obtained after the semantic information interaction.}
			\label{fig4}
			\vspace{-10pt}
		\end{figure}
		
		\subsection{PVT Encoder}
		To obtain the weights for the bi-temporal images in the same manner, we construct a siamese PVT network with weight sharing as the encoder. The pyramid structure is used to achieve multi-scale feature extraction, and the self-attention mechanism captures long-range dependencies. To effectively address the corresponding issue, we divide the multi-scale features into shallow features $T_{i1}$, $T_{i2}$ and deep features $T_{i3}$, $T_{i4}$. The shallow features contain information related to local texture, which will be used to construct the CGRR branch. The deep features contain global and high-level semantic information, which will be used to build the GCCM and SCEM modules, and the specific calculation formulas are as follows:
		
		\begin{equation}\label{eq:1}
		\begin{gathered}
		T_{ij}=PVT_{Encoder}\left(T_{i}\right),
		\end{gathered}
		\end{equation}
		where $PVT_{Encoder}$ refers to the encoder constructed by the siamese PVT network, $T_{i}$ represents the bi-temporal image inputs, and $T_{ij}$ represents the multi-scale feature maps obtained by the encoder for the bi-temporal images, where i refers to the time phase number, and j refers to the scale number.

		\subsection{GCCM}
		To incorporate change information into the original feature stage, we perform information interaction on the deep-layer features that contain global information. A direct application of the attention mechanism lacks specificity. Therefore, based on the characteristics of the task, we first apply channel attention and spatial attention to independently enhance the deep-layer features, allowing the model to focus on buildings. Then, we apply linear attention to enable the interaction of building changes between bi-temporal images, constructing high-level semantic associations between images from different time-space contexts. This process is aimed not only at building global cross-correlations but also at preparing for feature fusion in the decoding stage, allowing change information to interact across time.
		Specifically, we improved the linear angle attention mechanism and designed the GCCM tailored for the RSCD task. The multi-scale feature maps $T_{ij}$ are divided into Query $Q_{ij}$ , Key $K_{ij}$, and Value $V_{ij}$ based on the attention mechanism, and the specific calculation formulas are as follows:
		
		\begin{equation}\label{eq:2}
		\begin{aligned}
		Sim\left( Q,\ K \right) &=1-\frac{1}{\pi}\cdot \left( \frac{\pi}{2}-\arcsin \left( Q\cdot K^T \right) \right) \\
		&\approx \frac{1}{2}+\frac{1}{\pi}\cdot Q\cdot K^T,
		\end{aligned}
		\end{equation}	
		
		\begin{equation}\label{eq:3}
		\begin{aligned}
		H_{1j}&=Sim\left( Q_{1j},\ K_{2j} \right) \cdot V_{1j} \\
		&\approx \frac{1}{2}\cdot V_{1j}+\frac{1}{\pi}\cdot Q_{1j}\cdot \left(K_{2j}^T\cdot V_{1j} \right),
		\end{aligned}
		\end{equation}	
		
		\begin{equation}\label{eq:4}
		\begin{aligned}
		H_{2j}&=Sim\left( Q_{2j},\ K_{1j} \right) \cdot V_{2j} \\
		&\approx \frac{1}{2}\cdot V_{2j}+\frac{1}{\pi}\cdot Q_{2j}\cdot \left(K_{1j}^T\cdot V_{2j} \right),
		\end{aligned}
		\end{equation}	
		where $Sim(Q, K)$ represents the simplified calculation function of the linear angle attention, and $H_{ij}$ denotes the feature weight after the change information interaction.
		
		\subsection{CGRR Branch}
		To effectively address the issue of missing detection for special-colored building changes, we focus on extracting shallow features containing local texture information during the feature extraction process and use these features as inputs to the CGRR branch. Within the CGRR, feature extraction is performed through multiple stacked convolutional blocks, and a residual structure is applied to ensure the stability of network training, preventing the vanishing gradient problem and ensuring efficient gradient computation during training. Additionally, considering the impact of changes at different scales on detection accuracy, we have designed a multi-scale structure to ensure that features extracted at different scales can be effectively aligned with the PVT features. Finally, the output of the CGRR branch interacts with the channel-level concatenated features $C_{j}$, providing feedback information, allowing the model to focus on texture change areas at different scales and thereby enhancing the detection of special-colored building changes, and the specific calculation formulas are as follows:
		\begin{equation}\label{eq:5}
		\begin{gathered}
		D_{j}=Diff\left( T_{1j},\ T_{2j} \right),
		\end{gathered}
		\end{equation}	
		
		\begin{equation}\label{eq:6}
		\begin{gathered}
		C_{j}=Cat\left( T_{1j},\ T_{2j} \right),
		\end{gathered}
		\end{equation}	
		
		\begin{equation}\label{eq:7}
		\begin{gathered}
		C_{j}=C_{j}+R\left(MCL\left( T_{1j} \right)\right),
		\end{gathered}
		\end{equation}	
		where $D_{j}$ represents the differential features of the multi-scale feature maps, $C_{j}$ represents the concatenated features of the multi-scale feature maps, $MCL$ indicates the multi-convolutional layer stacking, and $R$ refers to the overall residual structure.
		\subsection{SCEM}
		To further enhance the fusion of feedback information from CGRR and strengthen the correlation between the high-level semantic features generated by GCCM, we designed the SCEM. This module first divides the multi-scale feature maps into two parts for processing. The first part interacts dynamically with the features through various convolution kernels and feature concatenation, enabling the capture of local changes at different receptive fields. The second part uses average pooling to extract global context information, ensuring the model can capture global semantics. Finally, the module fuses these two parts of features to enhance the semantic understanding capability, and the specific calculation formulas are as follows:
		\begin{equation}\label{eq:8}
		\begin{gathered}
		MC=Conv_{1\times 1}\left( \sum{Conv_{k\times k}}\left( C_j \right) \right) \ k=\left\{ 3,5,7 \right\},
		\end{gathered}
		\end{equation}
		\begin{equation}\label{eq:9}
		\begin{gathered}
		GC=\sigma \left( AVGPooling\left( Conv_{1\times 1}\left( C_j \right) \right) \right),
		\end{gathered}
		\end{equation}
		\begin{equation}\label{eq:10}
		\begin{gathered}
		C_{j}=C_{j}+MC\times GC,
		\end{gathered}
		\end{equation}
		where $\sigma$ represents sigmoid, $MC$ represents multiple convolutions, and $GC$ represents global context.
		\subsection{CFD}
		The cross-fusion decoder achieves effective fusion of bi-temporal image features through a cross-attention mechanism. This decoder performs feature interaction between the two branches, which is a critical step in the BCD task, ensuring that the model can comprehensively understand and integrate the features extracted by previous modules, thereby enabling accurate building change detection. After the feature fusion is completed, the features undergo upsampling processing through two convolutional blocks, which restores the spatial resolution of the image while ensuring that spatial detail information is not lost, effectively completing image reconstruction.
		
		\begin{figure*}
			\centering
			\includegraphics[width=1\textwidth]{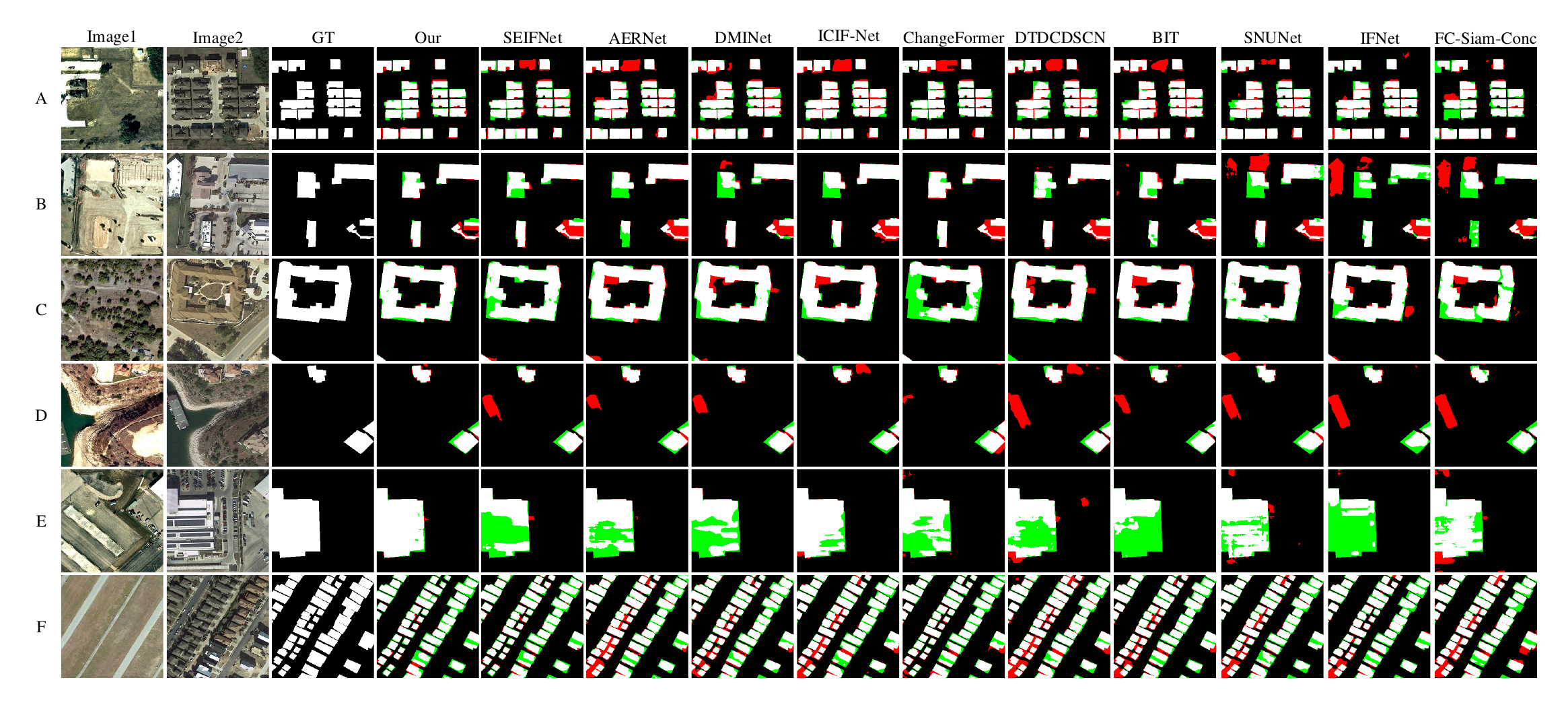}
			\caption{The comparison of the prediction results of CGCCE-Net with mainstream methods on the LEVIR-CD dataset. In this figure, the green areas represent the predicted parts that are missing compared to the GT, while the red areas represent the redundant predicted parts compared to GT.}
			\label{fig5}
			\vspace{-10pt}
		\end{figure*}

		\section{Experiment and evaluations}
		In this section, we will first introduce three public datasets: LEVIR-CD \cite{40}, WHU-CD \cite{41}, and GZ-CD \cite{42}. Next, we will describe the experimental environment, followed by an introduction to the evaluation metrics used in the experiments. Then, we will present the results of the comparative experiments and ablation studies, and finally, provide a comparison of parameter and computational complexity.
		
		\subsection{Datasets}
		We use three classical RSCD datasets to demonstrate the superiority of our method.
		
		\textbf{LEVIR-CD: }The LEVIR-CD dataset is specifically designed for the RSCD task, focusing on building change detection. It provides high-resolution optical remote sensing imagery, covering a variety of environments such as urban, rural, and architectural areas. Each image pair consists of two images taken at different time points and is accompanied by pixel-level change annotations that highlight the actual change areas of buildings. The dataset also includes various land cover types, complex backgrounds, and multi-scale changes, making it effective for evaluating the model's generalization ability and robustness in complex environments. Specifically, it contains 637 image pairs with a resolution of 1024×1024, and we reorganized them into $256\times256$ resolution image pairs by random cropping. The dataset is then split into training, testing, and validation sets with a 7:2:1 ratio.
		
		\textbf{WHU-CD: }The WHU-CD dataset is created by the Remote Sensing Institute of Wuhan University, focuses on BCD tasks in the context of urban expansion. This dataset includes high-resolution remote sensing image pairs from two different time points, primarily collected from urban areas. It covers changes in various land cover types, including buildings, roads, and green spaces. Each image pair provides pixel-level change annotations, clearly marking the regions that have changed as well as the unchanged parts. The dataset offers representative, real-world change data, including remote sensing images from urban areas with complex backgrounds and varying scales of change. It aims to enhance the robustness and adaptability of models in different land cover types and change patterns. We cropped the images into $256\times256$ resolution, resulting in a total of 7620 image pairs, which were then split into training, testing, and validation sets with an 8:1:1 ratio.
		
		\textbf{GZ-CD: }The GZ-CD dataset is specifically designed for RSCD tasks, focusing on building change detection. This dataset contains high-resolution remote sensing image pairs, covering various urban areas. Each pair of images is accompanied by pixel-level change annotations, accurately marking the changed regions and unchanged areas. The primary purpose of this dataset is to provide diverse training and testing data for models, including dense clusters of large buildings as well as small, isolated structures. It also includes complex scenarios, such as building shadows and overlapping building areas, making it suitable for evaluating models' accuracy in change detection. We cropped the images to a resolution of $256\times256$, using 2834, 325, and 400 image pairs for training, testing, and validation, respectively.
		
		\begin{table}[h]
			\caption{Indicator results for the LEVIR-CD dataset. Red color represents the best results and blue color represents the second best results (\%).\label{tab1}}
			\centering
			\renewcommand\arraystretch{1.2}
			\resizebox{0.49\textwidth}{!}
			{
				\begin{tabular}{ccccc}
					\hline
					\textbf{Methods}        & \textbf{F1} & \textbf{IoU} & \textbf{Precision} & \textbf{Recall} \bigstrut[t] \\
					\hline
					\textbf{FC-Siam-Conc\cite{25}}   & 81.77 & 69.16 & 84.17 & 79.49 \bigstrut[t] \\
					\textbf{IFNet       \cite{27}}   & 88.13 & 78.77 & 94.02 & 82.93 \bigstrut[t] \\
					\textbf{SNUNet      \cite{28}}   & 88.16 & 78.83 & 89.18 & 87.17 \bigstrut[t] \\
					\textbf{BIT         \cite{35}}   & 89.31 & 80.68 & 89.24 & 89.37 \bigstrut[t] \\
					\textbf{DTCDSCN     \cite{43}}   & 87.67 & 78.05 & 88.53 & 86.83 \bigstrut[t] \\
					\textbf{ChangeFormer\cite{44}}   & 90.40 & 82.48 & 92.05 & 88.80 \bigstrut[t] \\
					\textbf{ICIF-Net    \cite{45}}   & \textcolor{blue}{\textbf{91.18}} & \textcolor{blue}{\textbf{83.85}} & 91.13 & \textcolor{blue}{\textbf{90.57}} \bigstrut[t] \\
					\textbf{DMINet      \cite{46}}   & 90.71 & 82.99 & 92.52 & 89.95 \bigstrut[t] \\
					\textbf{AERNet      \cite{47}}   & 90.78 & 83.11 & 89.97 & \textcolor{red}{\textbf{91.59}} \bigstrut[t] \\
					\textbf{SEIFNet     \cite{48}}   & 90.86 & 83.25 & \textcolor{blue}{\textbf{94.29}} & 87.67 \bigstrut[t] \\
					\textbf{Ours                 }   & \textcolor{red}{\textbf{91.84}} & \textcolor{red}{\textbf{84.91}} & \textcolor{red}{\textbf{94.80}} & 89.05 \bigstrut[t] \\
					\hline
				\end{tabular}
			}
		\end{table}
		
		\subsection{Implementation Environment}
		We set up the code execution environment on the Ubuntu 18.04 operating system, enabling code debugging and experimental analysis. The model was implemented and trained under the PyTorch framework using an NVIDIA TITAN RTX 24GB GPU. During the experimental process, we determined a series of optimal parameters suitable for the model. The model was trained for a total of 500 epochs, with validation performed at the end of each epoch to save the best model. We used the AdamW optimizer and set the initial learning rate to 5e-4, with the cosine annealing algorithm employed for dynamic learning rate adjustment. The loss function was chosen as binary cross-entropy loss to guide the model's optimization, and its mathematical expression is as follows:
		\begin{equation}\label{eq:11}
		\begin{aligned}
		\operatorname{BCELoss}\left(y_{\text {i}}, \hat{y}_{\text {i}}\right)&=-\frac{1}{N} \sum_{i=1}^{N}\left[y_{\text {i}} \log \left(\hat{y}_{\text {i}}\right)	\right.	\\
		&\left.+\left(1-y_{\text {i}}\right) \log \left(1-\hat{y}_{\text {i}}\right)\right],
		\end{aligned}
		\end{equation}
		where N denotes the number of pixels, $y_{i}$ represents the true value, and $\hat{y}_{i}$ denotes the predicted value.
		
		\begin{figure*}
			\centering
			\includegraphics[width=1\textwidth]{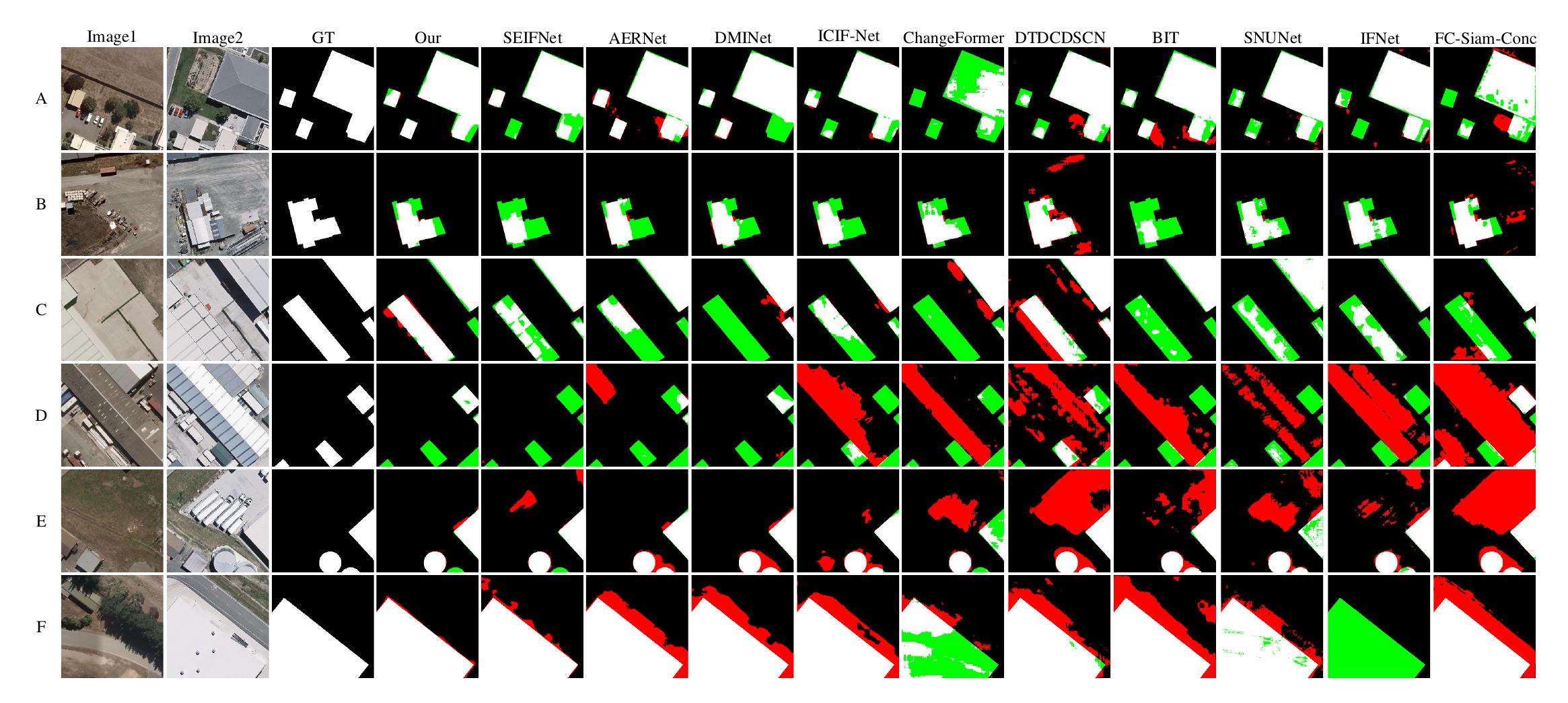}
			\caption{The comparison of the prediction results of CGCCE-Net with mainstream methods on the WHU-CD dataset. In this figure, the green areas represent the predicted parts that are missing compared to the GT, while the red areas represent the redundant predicted parts compared to GT.}
			\label{fig6}
			\vspace{-10pt}
		\end{figure*}
		
		\begin{table}[h]
			\caption{Indicator results for the WHU-CD dataset. Red color represents the best results and blue color represents the second best results (\%).\label{tab2}}
			\centering
			\renewcommand\arraystretch{1.2}
			\resizebox{0.49\textwidth}{!}
			{
				\begin{tabular}{ccccc}
					\hline
					\textbf{Methods}        & \textbf{F1} & \textbf{IoU} & \textbf{Precision} & \textbf{Recall} \bigstrut[t] \\
					\hline
					\textbf{FC-Siam-Conc\cite{25}}   & 72.61 & 56.99 & 75.89 & 69.30 \bigstrut[t] \\
					\textbf{IFNet       \cite{27}}   & 83.40 & 71.52 & \textcolor{red}{\textbf{96.91}} & 73.19 \bigstrut[t] \\
					\textbf{SNUNet      \cite{28}}   & 88.34 & 79.11 & 91.34 & 85.53 \bigstrut[t] \\
					\textbf{BIT         \cite{35}}   & 87.47 & 77.73 & 88.71 & 86.27 \bigstrut[t] \\
					\textbf{DTCDSCN     \cite{43}}   & 90.48 & 82.62 & 91.84 & 89.16 \bigstrut[t] \\
					\textbf{ChangeFormer\cite{44}}   & 86.88 & 76.81 & 88.50 & 85.33 \bigstrut[t] \\
					\textbf{ICIF-Net    \cite{45}}   & 90.77 & 83.09 & 92.93 & 88.70 \bigstrut[t] \\
					\textbf{DMINet      \cite{46}}   & 91.49 & 84.31 & 92.65 & 90.35 \bigstrut[t] \\
					\textbf{AERNet      \cite{47}}   & 92.18 & 85.49 & 92.47 & 91.89 \bigstrut[t] \\
					\textbf{SEIFNet     \cite{48}}   & \textcolor{blue}{\textbf{93.29}} & \textcolor{blue}{\textbf{87.43}} & 93.99 & \textcolor{blue}{\textbf{92.61}} \bigstrut[t] \\
					\textbf{Ours                }    & \textcolor{red}{\textbf{94.90}} & \textcolor{red}{\textbf{90.29}} & \textcolor{blue}{\textbf{96.59}} & \textcolor{red}{\textbf{93.26}} \bigstrut[t] \\
					\hline
				\end{tabular}
			}
		\end{table}
		
		\subsection{Evaluation Metrics}
		We use four common evaluation metrics: F1, IoU, Precision, and Recall, to compare the performance of our model with other BCD methods. F1 combines Precision and Recall, making it especially useful for handling class imbalance, and provides a comprehensive performance evaluation. IoU measures the overlap between the predicted results and the ground truth labels, thus evaluating the accuracy of change detection. Precision measures the proportion of correctly identified positive samples, while Recall assesses the model's ability to identify all positive samples. The specific mathematical formulas for these metrics are as follows:
		\begin{equation}\label{eq:12}
		\begin{gathered}
		F1=\frac{2TP}{2TP+FP+FN},
		\end{gathered}
		\end{equation}
		\begin{equation}\label{eq:13}
		\begin{gathered}
		IoU=\frac{TP}{TP+FP+FN},
		\end{gathered}
		\end{equation}
		\begin{equation}\label{eq:14}
		\begin{gathered}
		Precision=\frac{TP}{TP+FP},
		\end{gathered}
		\end{equation}
		\begin{equation}\label{eq:15}
		\begin{gathered}
		Recall=\frac{TP}{TP+FN},
		\end{gathered}
		\end{equation}
		where TP represents true positives, FP represents false positives, FN represents false negatives, and TN represents true negatives.
		
		\begin{figure*}
			\centering
			\includegraphics[width=1\textwidth]{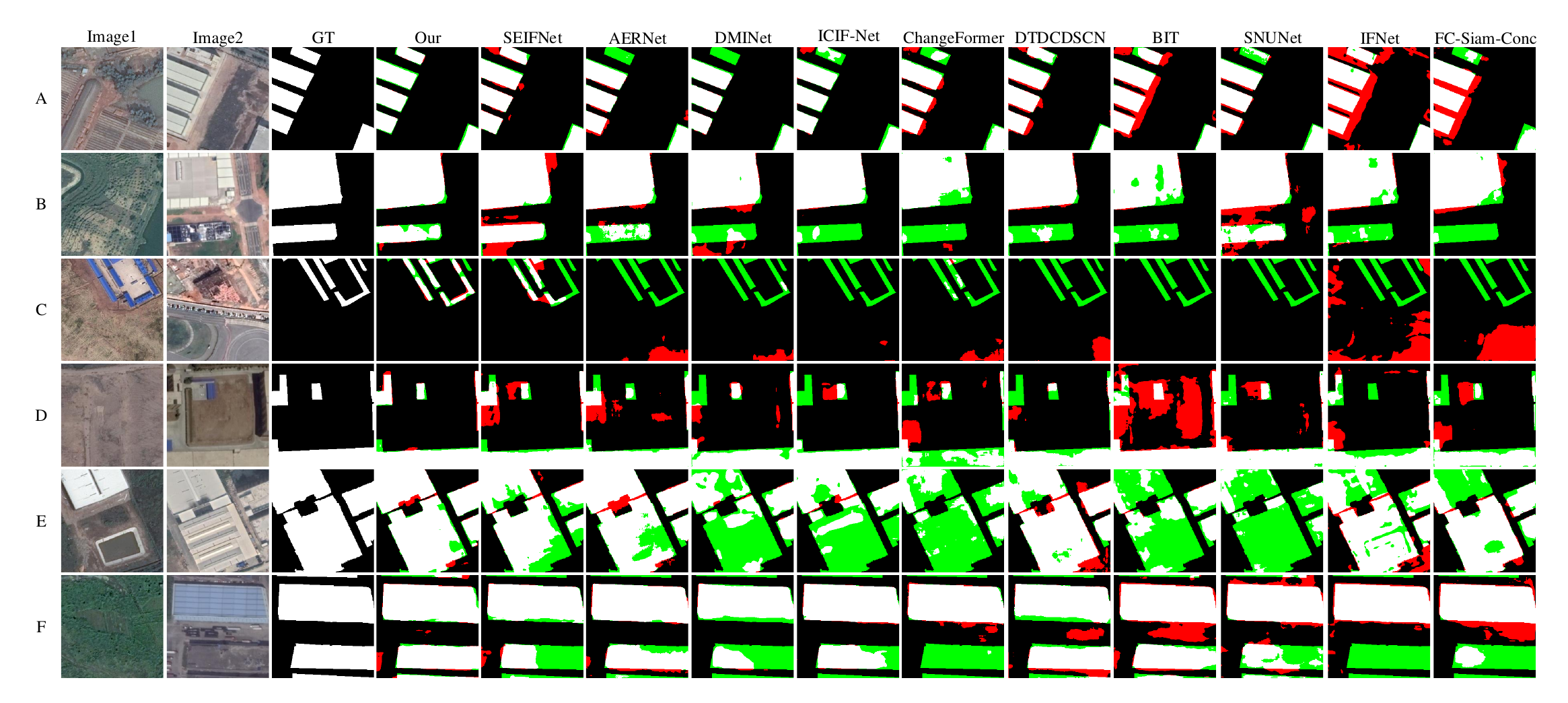}
			\caption{The comparison of the prediction results of CGCCE-Net with mainstream methods on the GZ-CD dataset. In this figure, the green areas represent the predicted parts that are missing compared to the GT, while the red areas represent the redundant predicted parts compared to GT.
			}
			\label{fig7}
			\vspace{-10pt}
		\end{figure*}

		\subsection{Comparative Experiment}
		In this section, We compare our CGCCE-Net with several classic and mainstream BCD methods for the RSCD task based on multiple data metrics and representative visual prediction results.
		
		FC-Siam-Conc constructs an architecture suitable for BCD tasks based on a U-shape structure, and experiments with concatenation, difference, and fusion approaches, laying the foundation for the diversification of subsequent BCD methods. 
		
		IFNet combines multi-scale features with attention modules to achieve deep feature extraction and fuses deep features with difference features, guiding model optimization through deep supervision. 
		
		SNUNet implements deep cross-scale feature interaction through multi-level dense connections, further enhancing semantic information with channel attention. 
		
		BIT proposes using a Transformer architecture as an encoder for feature extraction and global context modeling, effectively capturing long-range dependencies to optimize feature information. 
		
		DTCDSCN \cite{43} proposes performing change detection and building extraction tasks simultaneously, introducing multiple attention mechanisms to capture dependencies between channels and spatial dimensions, and improving the loss function to address sample imbalance. 
		
		ChangFormer \cite{44} combines the hierarchical structure of Transformer with MLP decoders to achieve multi-scale long-range detail extraction. 
		
		ICIF-Net \cite{45} integrates CNN and Transformer architectures to explore the potential advantages of combining local and global information and applies a linear convolutional attention module to assist the organic fusion of the two architectures. 
		
		DMINet \cite{46} introduces a dual-branch architecture to implement multi-level difference aggregation and applies a joint attention block to guide global feature distribution, enhancing inter-layer information interaction. 
		
		AERNet \cite{47} proposes global context feature aggregation and enhanced coordinate attention to aggregate multi-layer context information, constructing channel and spatial dependencies in feature maps, while designing an adaptive weighted loss function combined with deep supervision to achieve edge refinement. 
		
		SEIFNet \cite{48} introduces a spatiotemporal difference enhancement module to capture multi-level global and local information and designs an adaptive context fusion module to build a progressive decoder, achieving inter-layer feature information interaction guided by semantic information.
		
		\begin{table}[h]
			\caption{Indicator results for the GZ-CD dataset. Red color represents the best results and blue color represents the second best results (\%).\label{tab3}}
			\centering
			\renewcommand\arraystretch{1.2}
			\resizebox{0.49\textwidth}{!}
			{
				\begin{tabular}{ccccc}
					\hline
					\textbf{Methods}        & \textbf{F1} & \textbf{IoU} & \textbf{Precision} & \textbf{Recall} \bigstrut[t] \\
					\hline
					\textbf{FC-Siam-Conc\cite{25}}   & 74.23 & 59.03 & 80.37 & 68.97 \bigstrut[t] \\
					\textbf{IFNet       \cite{27}}   & 82.15 & 69.71 & \textcolor{blue}{\textbf{92.19}} & 74.08 \bigstrut[t] \\
					\textbf{SNUNet      \cite{28}}   & 84.25 & 72.79 & 84.25 & 81.82 \bigstrut[t] \\
					\textbf{BIT         \cite{35}}   & 80.23 & 66.99 & 82.40 & 78.18 \bigstrut[t] \\
					\textbf{DTCDSCN     \cite{43}}   & 83.00 & 70.93 & 88.19 & 78.38 \bigstrut[t] \\
					\textbf{ChangeFormer\cite{44}}   & 73.66 & 58.30 & 84.59 & 65.23 \bigstrut[t] \\
					\textbf{ICIF-Net    \cite{45}}   & 85.09 & 74.05 & 89.90 & 80.76 \bigstrut[t] \\
					\textbf{DMINet      \cite{46}}   & 81.98 & 69.46 & 87.92 & 76.79 \bigstrut[t] \\
					\textbf{AERNet      \cite{47}}   & 84.42 & 73.03 & 88.06 & 81.07 \bigstrut[t] \\
					\textbf{SEIFNet     \cite{48}}   & \textcolor{blue}{\textbf{87.48}} & \textcolor{blue}{\textbf{77.75}} & 89.64 & \textcolor{blue}{\textbf{85.43}} \bigstrut[t] \\
					\textbf{Ours                }    & \textcolor{red}{\textbf{89.90}} & \textcolor{red}{\textbf{81.65}} & \textcolor{red}{\textbf{94.16}}  & \textcolor{red}{\textbf{86.00}} \bigstrut[t] \\
					\hline
				\end{tabular}
			}
		\end{table}
		
		\begin{figure*}
			\centering
			\includegraphics[width=1\textwidth]{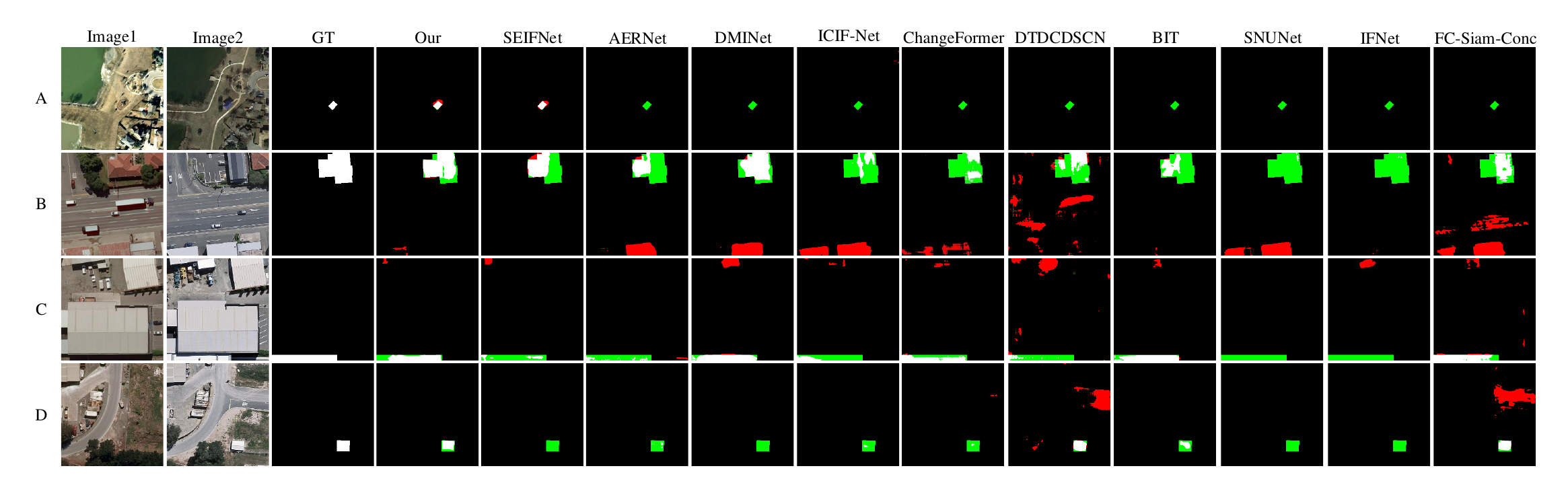}
			\caption{The comparison of CGCCE-Net with mainstream methods in detecting changes in buildings with special colors is shown in the figure. In this image, the green areas represent the predicted regions that are missing compared to the GT, while the red areas represent the redundant predictions compared to the GT.}
			\label{fig8}
			\vspace{-10pt}
		\end{figure*}

		As shown in Fig. \ref{fig5}, we selected representative samples A-F from the LEVIR-CD dataset and compared the prediction maps generated by CGCCE-Net with the results of other methods. In samples A and F, predictions were made for densely packed small buildings with similar shapes. Specifically, in sample A, other methods overfitted and mistakenly identified construction sites as change areas, while in sample F, other methods failed to accurately detect the buildings, predicting two rows of buildings as a single row. Samples B, C, and D involve predictions for fewer buildings. Our method accurately predicted the main differences, with neither large prediction omissions nor excessive redundancies. Sample E focuses on buildings with alternating shadow and white colors. Other methods struggled to distinguish the relationship between the buildings and their shadows, whereas our method made a complete prediction without any strip-shaped omissions. Facing these three different types of prediction issues, our model demonstrated better generalization capability, achieving precise building detection and change prediction. It also deepened the model's understanding of semantic information, successfully overcoming the problem of prediction omissions for buildings with special color patterns like shadow and white alternating buildings. As shown in Table. \ref{tab1}, the corresponding metrics for the visual results of each method on the LEVIR-CD dataset clearly display that our method outperforms the second-place SEIFNet by 0.98\% in F1 and 1.66\% in IoU.

		\begin{figure*}
			\centering
			\includegraphics[width=1\textwidth]{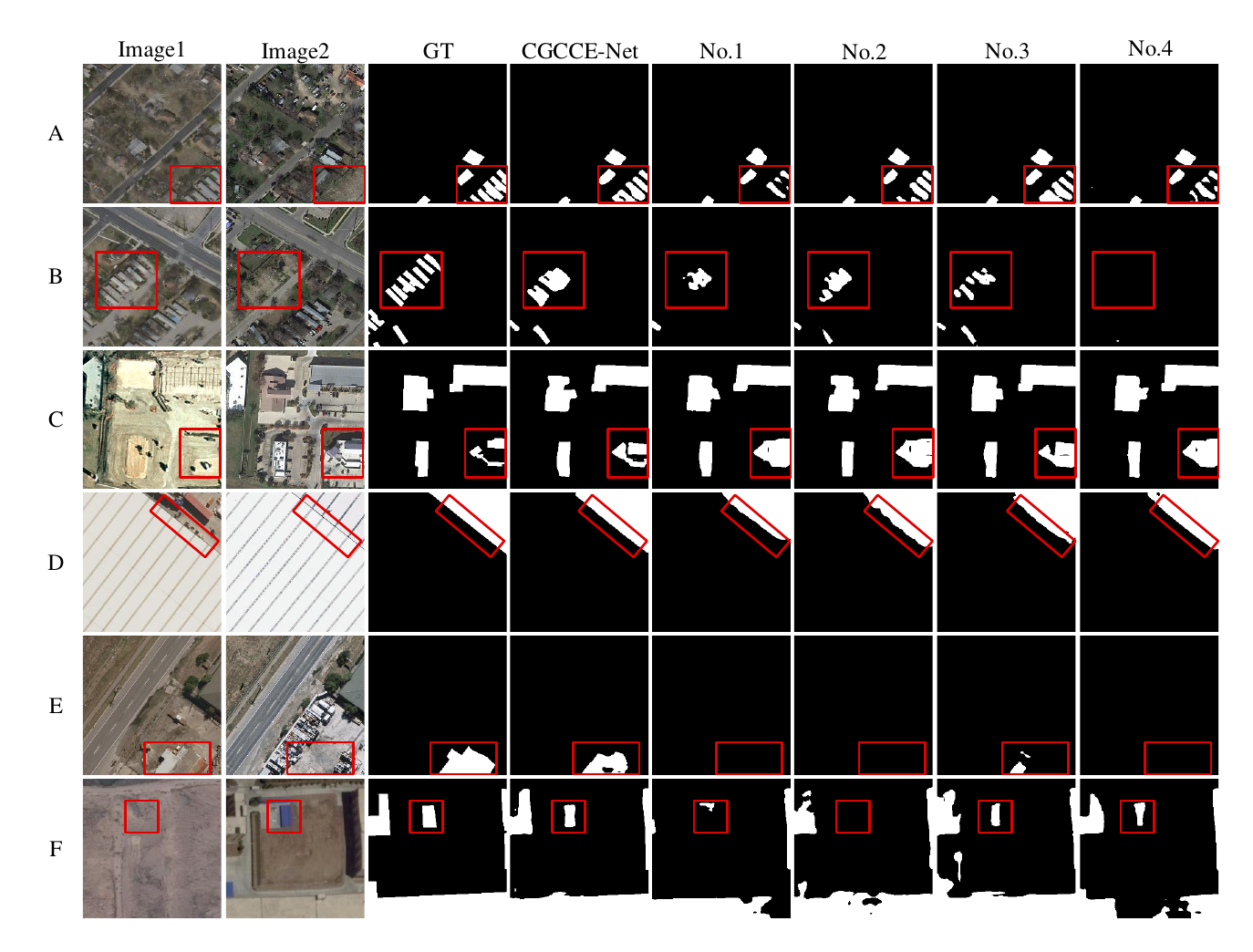}
			\caption{The ablation experiment of CGCCE-Net on three datasets, with the key focus areas in the samples highlighted using red boxes.}
			\label{fig9}
			\vspace{-10pt}
		\end{figure*}
	
		\begin{table*}[!ht]
			\centering
			\setlength{\tabcolsep}{11.8pt}
			\caption{\textrm{Overview of ablation results for CGCCE-Net and metric outcomes on three datasets (\%).}}
			\begin{tabular}{ccccccccccc}
				\hline
				\multirow{2}{*}{\textbf{No.}} & \multirow{2}{*}{\textbf{GCCM}} & \multirow{2}{*}{\textbf{CGRR}} & \multirow{2}{*}{\textbf{SCEM}} & \multirow{2}{*}{CFD} & \multicolumn{2}{c}{\textbf{LEVIR-CD}} & \multicolumn{2}{c}{\textbf{WHU-CD}} & \multicolumn{2}{c}{\textbf{GZ-CD}} \bigstrut[t] \\
				\cline{6-11}
				& & & & & \textbf{F1} & \textbf{IoU} & \textbf{F1} & \textbf{IoU} & \textbf{F1} & \textbf{IoU} \bigstrut[t] \\
				\hline
				\textbf{1}   & $\times$ & \checkmark & \checkmark & \checkmark
				& 91.52 & 84.36 & 94.36 & 89.32 & 89.43 & 80.88 \bigstrut[t] \\
				\textbf{2}   & \checkmark & $\times$ & \checkmark & \checkmark
				& 91.50 & 84.34 & 94.44 & 89.46 & 89.39 & 80.81 \bigstrut[t] \\
				\textbf{3}   & \checkmark & \checkmark & $\times$ & \checkmark
				& 91.61 & 84.51 & 94.67 & 89.87 & 89.52 & 81.03 \bigstrut[t] \\
				\textbf{4}   & \checkmark & \checkmark & \checkmark & $\times$
				& 91.64 & 84.57 & 94.47 & 89.52 & 89.56 & 81.10 \bigstrut[t] \\
				\hline
			\end{tabular}
			\label{tab4}
		\end{table*}
		
		\begin{table*}
			\centering
			\setlength{\tabcolsep}{13pt}
			\captionsetup{justification=raggedright, singlelinecheck=false}
			\caption{Comparison of Parameters$\left( M\right)$ and FLOPs$\left( G\right)$, along with a brief description of the comparison methods for F1(\%) and IoU(\%) performance.}
			\begin{tabular}{cccccccccc}
				\hline
				\multirow{2}{*}{\textbf{Methods}} & \multirow{2}{*}{\textbf{Params}} & \multirow{2}{*}{\textbf{FLOPs}} & \multicolumn{2}{c}{\textbf{LEVIR-CD}} & \multicolumn{2}{c}{\textbf{WHU-CD}} & \multicolumn{2}{c}{\textbf{GZ-CD}} \bigstrut[t] \\
				\cline{4-9}
				& & & \textbf{F1} & \textbf{IoU} & \textbf{F1} & \textbf{IoU} & \textbf{F1} & \textbf{IoU} \bigstrut[t] \\
				\hline
				\textbf{FC-Siam-Conc\cite{25}}   & 1.55  & 5.32   & 81.77 & 69.16 & 72.61 & 56.99 & 74.23 & 59.03 \bigstrut[t] \\
				\textbf{IFNet\cite{27}}          & 50.71 & 82.35  & 88.13 & 78.77 & 83.40 & 71.52 & 82.15 & 69.71 \bigstrut[t] \\
				\textbf{SNUNet\cite{28}}         & 12.03 & 54.88  & 88.16 & 78.83 & 88.34 & 79.11 & 84.25 & 72.79 \bigstrut[t] \\
				\textbf{BIT\cite{35}}            & 3.55  & 10.59  & 89.31 & 80.68 & 87.47 & 77.73 & 80.23 & 66.99 \bigstrut[t] \\
				\textbf{DTCDSCN\cite{43}}        & 41.07 & 13.21  & 87.67 & 78.05 & 90.48 & 82.62 & 83.00 & 70.93 \bigstrut[t] \\
				\textbf{ChangeFormer\cite{44}}   & 41.03 & 202.83 & 90.40 & 82.48 & 86.88 & 76.81 & 73.66 & 58.30 \bigstrut[t] \\
				\textbf{ICIF-Net\cite{45}}       & 25.83 & 25.27  & 91.18 & 83.85 & 90.77 & 83.09 & 85.09 & 74.05 \bigstrut[t] \\
				\textbf{DMINet\cite{46}}         & 6.24  & 14.55  & 90.71 & 82.99 & 91.49 & 84.31 & 81.98 & 69.46 \bigstrut[t] \\
				\textbf{AERNet\cite{47}}         & 25.36 & 12.82  & 90.78 & 83.11 & 92.18 & 85.49 & 84.42 & 73.03 \bigstrut[t] \\
				\textbf{SEIFNet\cite{48}}        & 8.37  & 27.9   & 90.86 & 83.25 & 93.68 & 88.11 & 87.48 & 77.75 \bigstrut[t] \\
				\textbf{Ours}                    & 56.67 & 17.57  & \textbf{91.84} & \textbf{84.91} & \textbf{94.90} & \textbf{90.29} & \textbf{89.90} & \textbf{81.65} \bigstrut[t] \\
				\hline
			\end{tabular}
			\label{tab5}
		\end{table*}

		\begin{figure}[!t]
			\centering
			\includegraphics[width=3.5in]{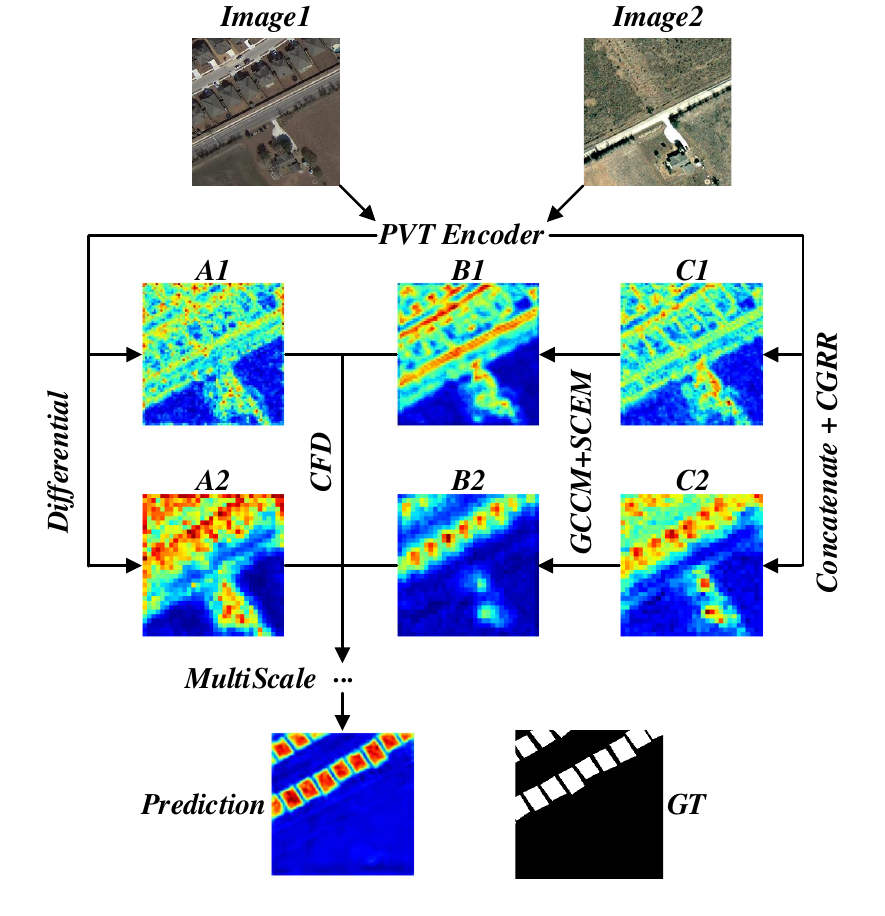}
			\caption{The simplified CGCCE-Net architecture diagram intuitively demonstrates the functions and roles of each module at different scales, providing a visual effect.}
			\label{fig10}
			\vspace{-10pt}
		\end{figure}

		As shown in Fig. \ref{fig6}, we selected representative samples A-F from the WHU-CD dataset and compared the prediction maps generated by CGCCE-Net with the results of other methods. In sample B, the building changes are divided into three levels based on their color differences with the background: distinctly visible, somewhat visible, and nearly identical colors. SEIFNet, AERNet, and several other mainstream methods can only distinguish changes in buildings with a distinctly visible color difference, while our method clearly identifies the difference in regions with less obvious color changes. In sample C, the color of the difference region undergoes a generally similar change. Among various methods, only ours can accurately predict the entire building in the lower-left corner, reflecting the model’s deep understanding of semantic information and its ability to achieve cross-temporal semantic cognition with bi-temporal images. In sample F, large-scale difference areas can be completely predicted by both classical and mainstream methods. However, the appearance of a road caused many methods to mistakenly identify it as a change in the building, leading to significant redundant predictions. Only SEIFNet can somewhat distinguish the road, while our method almost completely avoids redundant predictions. As shown in Table. \ref{tab2}, the corresponding metrics for the visual results of each method on the WHU-CD dataset clearly indicate that our method outperforms SEIFNet by 1.61\% in F1 and 2.86\% in IoU.

		As shown in Fig. \ref{fig7}, we selected representative samples A-F from the GZ-CD dataset and compared the prediction maps generated by CGCCE-Net with the results of other methods. In sample A, AERNet and DMINet successfully predicted the three difference regions with similar colors, but they failed to predict the upper difference region with a larger color and scale difference. SEIFNet, on the other hand, showed redundant predictions of false changes in the upper-left corner. Our method, through early change information interaction using the CGRR branch and GCCM, enhances the model's ability to perceive changes and distinguish false changes, solving the issues encountered by other methods. In sample C, our method predicted the irregular, fine strip-shaped change areas more completely, reflecting the model's strong generalization ability. Overall, in samples B, C, and D, most of the change regions are deep blue. Apart from our method and SEIFNet, other methods failed to solve the problem of detecting changes in buildings with special colors. Compared to SEIFNet, our method provides higher accuracy, stronger edge refinement capability, and better false change distinction. As shown in Table. \ref{tab3}, the corresponding metrics for the visual results of each method on the GZ-CD dataset clearly indicate that our method outperforms SEIFNet by 2.42\% in F1 and 3.9\% in IoU.

		Overall, the representative samples selected from multiple datasets visually demonstrate the superiority of CGCCE-Net from different perspectives, including scenarios with similar background colors, dense or sparse building distributions, and irregular difference areas. Most importantly, our method effectively solves the problem of detecting changes in buildings with special colors. As shown in Fig. \ref{fig8}, we have once again selected four sets of visualized images from multiple datasets, which represent the predictive results of changes in buildings with special colors, highlighting CGCCE-Net's ability to address this specific issue.

		\subsection{Ablation Study}
		
		\begin{figure*}
			\centering
			\includegraphics[width=1\textwidth]{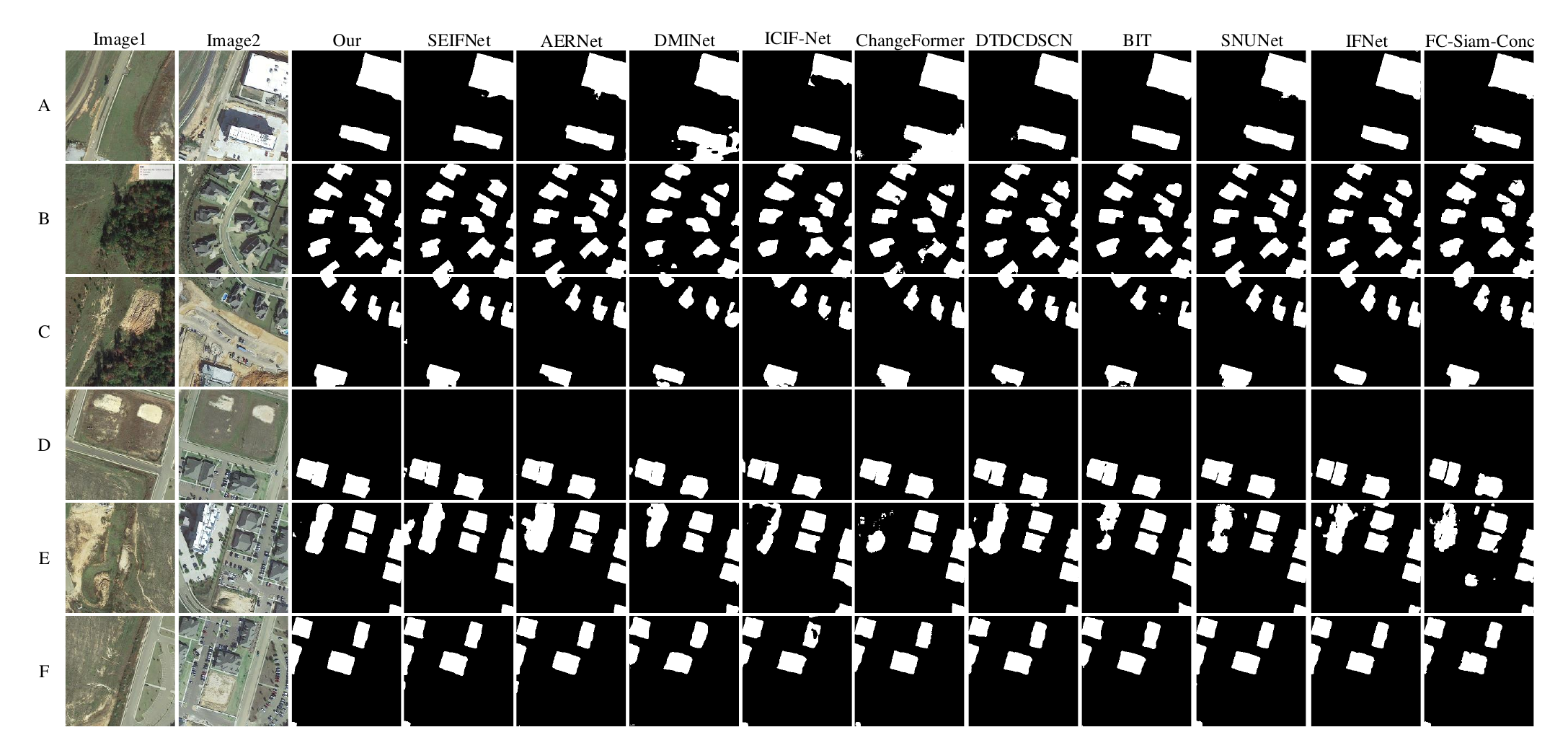}
			\caption{The comparison of CGCCE-Net with mainstream methods in predicting building change detection results in real-world scenarios is shown. This scenario uses Google Earth images of Oxford, Mississippi (2012-2019).}
			\label{fig11}
			\vspace{-10pt}
		\end{figure*}
		
		\begin{figure}[!t]
			\centering
			\includegraphics[width=3.4in]{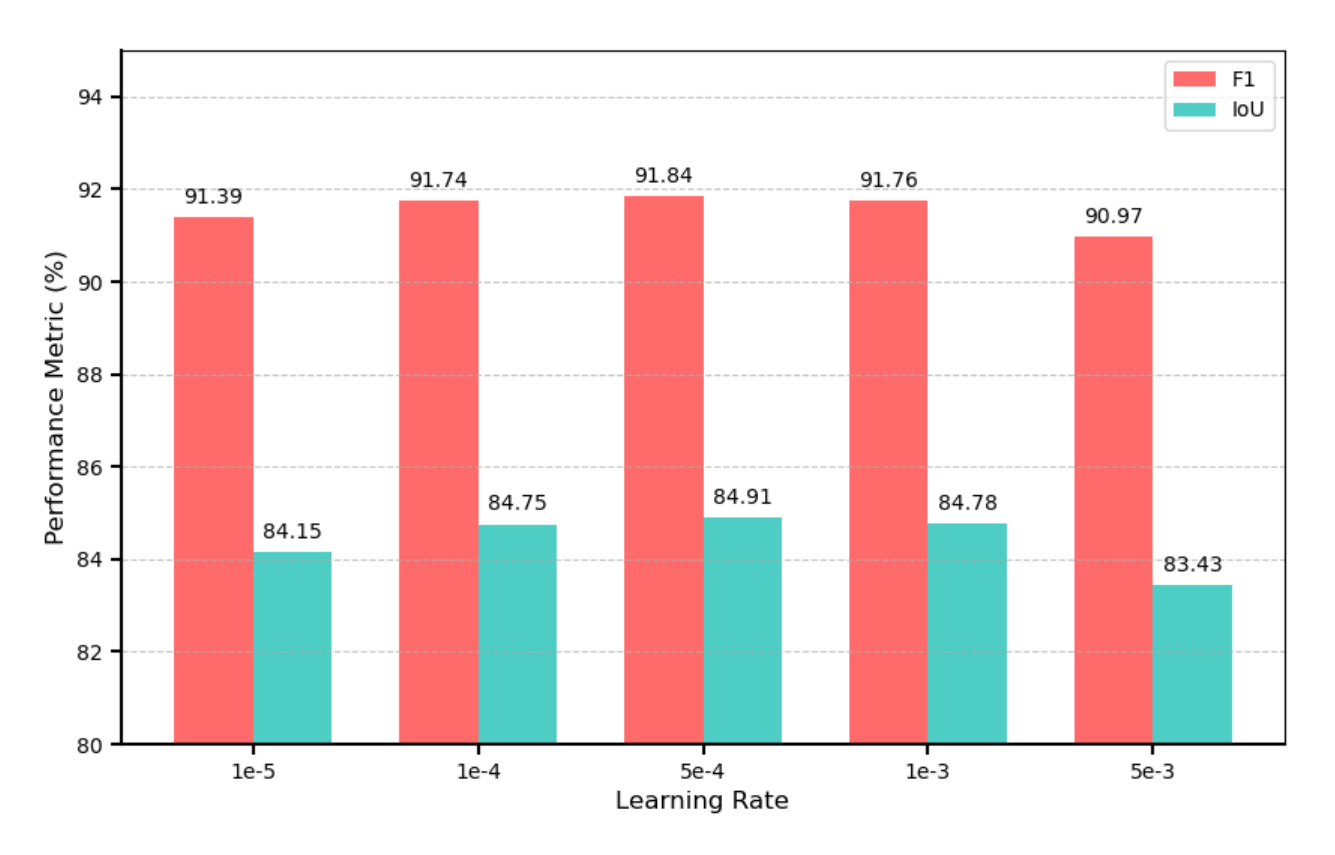}
			\caption{The sensitivity analysis results of key parameters in CGCCE-Net are presented, including the F1 and IoU metrics of the model under different learning rates.}
			\label{fig12}
			\vspace{-10pt}
		\end{figure}
		
		\begin{figure}[!t]
			\centering
			\includegraphics[width=3.4in]{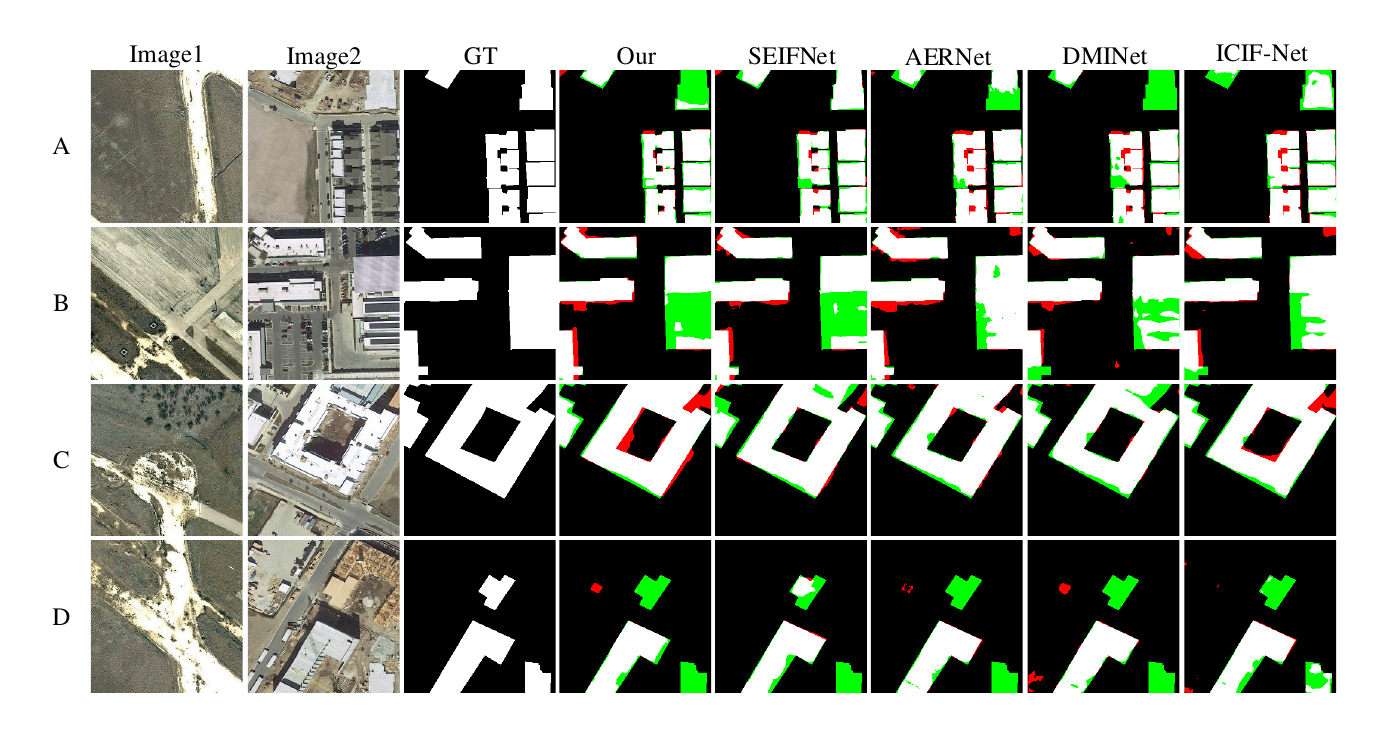}
			\caption{The results illustrate that CGCCE-Net still exhibits instances of missing target predictions and redundant predictions.}
			\label{fig13}
			\vspace{-10pt}
		\end{figure}
		
		As shown in Table. \ref{tab4}, we conducted an ablation experiment on the four components of the CGCCE-Net model to verify its effectiveness. The model with the GCCM removed is referred to as No. 1, the model with the CGRR branch removed is referred to as No. 2, the model with the SCEM removed is referred to as No. 3, and the model with the CFD removed is referred to as No. 4. The ablation experiment results of CGCCE-Net on three datasets are clearly presented.

		As shown in Fig. \ref{fig9}, we selected representative samples from multiple datasets and highlighted the key areas with red boxes for clearer presentation. First, a horizontal observation reveals that the overall prediction results of CGCCE-Net are significantly better than those of the ablated models, with predictions closer to the GT in terms of main objects and edge details. Vertically, by observing the results of No. 1, it can be seen that GCCM enhances global semantic information and enables cross-temporal and spatial interactions between bi-temporal images. After removing this part in No. 1, effective detection of the difference areas becomes difficult, especially in samples A, B, E, and F, where parts of the dense small buildings are missing, intertwined, and even some entire buildings fail to be detected. The performance of No. 1 and No. 2 in Sample F highlights the significance and advancement of our approach when dealing with building change detection in special colors. The CGRR branch extracts texture-based feature information from shallow features and guides the model to address this specific problem, while GCCM enables cross-temporal and spatial semantic information interaction in deep features. The combination of these two parts ultimately achieves the target.
		
		As shown in Fig. \ref{fig10}, we selected some visualized images to present a more intuitive training process. Image1 and Image2 serve as the input of bi-temporal images, which are independently trained through the PVT encoder. Afterward, differential operations are performed at multiple scales, and the stitching operation guided by early change information based on the CGRR branch generates images Aj and Cj. These are then processed through the semantic information interaction and global information enhancement provided by the GCCM and SCEM. Notably, the process from C1 to B1 clearly demonstrates the enhancement of change information, while the process from C2 to B2 shows the suppression of change areas. Finally, through CFD, Aj and Bj are fused, and image reconstruction is performed to obtain the predicted image.

		To further demonstrate the effectiveness of the BCD method in real-world applications, we used image tiles from a real remote sensing scene to validate the model's effectiveness in handling random and complex scenarios. The visualized images are shown in Fig. \ref{fig11}. Specifically, we used the model weights trained with various BCD methods on the LEVIR-CD dataset to test the real-world scene. Although there is no GT reference for the prediction of this remote sensing image, it is still visually apparent that our model produces more structured and organized building segmentation, with a clear advantage in edge handling. This also validates the effectiveness of our method in addressing complex change scenarios.

		The sensitivity analysis of key model parameters is also crucial. Therefore, we conducted a multi-level division of the learning rate settings, as shown in Fig. \ref{fig12}, and performed analysis and validation on the independent LEVIR-CD dataset. Based on the F1 and IoU metrics, we observed that when the learning rate is between 1e-4 and 1e-3, the model training results tend to stabilize, reaching the optimal performance at 5e-4. At a learning rate of 1e-5, the model converges too slowly and falls into a local optimum in the later stages, failing to reach an optimal performance level. Conversely, at a learning rate of 5e-3, the model exhibits instability in the early training stages, frequently experiencing large fluctuations and skipping the optimal solution, ultimately failing to converge.

		\subsection{Parameters and FLOPs Comparison}
		As shown in Table. \ref{tab5}, we compare the parameter and FLOPs of CGCCE-Net with other methods. The parameter refers to the total number of trainable parameters in the model, including weights and biases, which impact storage requirements and computational cost during training. FLOPs measure the number of floating-point operations required for a single forward pass, directly affecting the network’s computational speed and efficiency. CGCCE-Net adopts a siamese network structure based on the PVT encoder and leverages the CGRR branch to guide early change information, resulting in a larger parameter count. However, when considering the ratio of parameter quantity to computational cost, our network achieves lower computational complexity while maintaining superior performance.
		
		\section{Limitations and Future Work}
		We specifically designed CGCCE-Net to address the challenge of building change detection with special color variations. After extensive experiments and analyses, the effectiveness of the network has been validated, achieving excellent performance. The CGRR branch is designed to extract local texture information, but while it provides guided change information, it also introduces local noise interference and increases computational overhead. The GCCM is designed to facilitate semantic information interaction between bi-temporal images, but the negative impact of noise on global feature representations during the absorption of long-range dependencies still requires further exploration. Additionally, the model lacks a more refined transition mechanism and an effective fusion strategy for change information. These issues result in certain limitations for the model. As shown in Fig. \ref{fig13}, CGCCE-Net still exhibits a significant degree of missing target predictions and redundant predictions in specific scenarios. Moreover, our work focuses more on solving the specific problem of building change detection, while to some extent overlooking the change detection in more diverse scenarios. Therefore, in future work, we will further explore balanced solutions for the inherent and specific challenges in RSCD, while ensuring the effective resolution of specific issues, and further enhancing the model's performance in diverse scenarios.
		
		\section{Conclusion}
		In this paper, we use a PVT-based encoder as a siamese network and construct CGCCE-Net to address the issue of building change detection with special color variations. Specifically, the model uses the PVT encoder to extract multi-scale feature information, which is then processed by pixel-level differential fusion and channel-level concatenation fusion, dividing it into two branches that contain different change information. We designed the CGRR branch to input shallow features and feedback channel-level concatenation information, enabling multi-level change information guidance based on texture features. In the deeper high-level features, we implement independent information enhancement using CAB and SAB, and then the GCCM utilizes linear attention to facilitate semantic information interaction between bi-temporal images, establishing the cognitive relationship between buildings and change targets in different images. The enhanced channel-level information is then processed through SCEM, where different convolution kernels and global context information are used, with a residual structure jointly achieving weight reallocation, strengthening the long-range dependencies of the independent feature maps and the semantic recognition relationships of the bi-temporal feature maps. Finally, the feature information from the two branches is fused using CFD to integrate change information, identify the target region, and perform image reconstruction. CGCCE-Net undergoes comparative and ablation experiments on multiple public datasets to verify the effectiveness and superiority of our method compared to mainstream BCD methods.

		
		\bibliographystyle{unsrt}
		
		\bibliography{reference.bib}

%
%

	\end{sloppypar}
\end{document}